\documentclass[pdflatex,sn-mathphys-num]{sn-jnl}

\usepackage{array}
\usepackage{graphicx}
\usepackage{multirow}
\usepackage{booktabs}
\usepackage{amsmath,amssymb,amsfonts}
\usepackage{mathtools}
\usepackage{subcaption}
\usepackage{hyperref}
\usepackage{enumitem}
\usepackage{float}
\usepackage{placeins}
\usepackage{soul,xcolor}

\newcommand{\NA}{\ensuremath{\mathrm{N/A}}}

\begin{document}

\title[RC-aware nnU-Netv2]{A Simple yet Effective Framework for Pre-treatment and Post-treatment Glioma Segmentation Using Multimodal MRI}

%% ----------------------------
%% Authors (please fill in)
%% ----------------------------
\author[1]{Lin Qu}
\author*[1]{Ziqi Chen}\email{zqchen@fem.ecnu.edu.cn}
\author[1]{Anqi Wu}
\author[1]{Jinyao Shen}
\author*[2]{Hai Shu}\email{hs120@nyu.edu}

\affil[1]{School of Statistics, East China Normal University, Shanghai, China}
\affil[2]{Department of Biostatistics, School of Global Public Health, New York University, New
York, NY, USA}

%% ----------------------------
%% Abstract & Keywords
%% ----------------------------
\abstract{
\textbf{Background:} Accurate glioma segmentation across pre- and post-treatment MRI remains challenging because treatment alters both image appearance and the target regions. In particular, the resection cavity (RC) is defined only
for post-treatment cases, complicating  joint learning from pooled data. 

\textbf{Methods:} We present RC-aware nnU-Netv2, a simple yet effective glioma segmentation framework that  combines lesion- and boundary-aware single-cohort training, an RC-aware joint  learning objective, and treatment-status-guided model routing at inference.
During pooled training, the joint objective applies both standard  and boundary-weighted binary cross-entropy to all four region channels, 
while employing Dice supervision for channels with non-empty targets and a controlled false-positive penalty for channels with empty targets across the mini-batch. 
This design explicitly accommodates the absence of RC in pre-treatment cases. Our approach   uses only the standard nnU-Netv2 data augmentation  and a single specialized model per case, without additional synthetic tumor generation or multi-model ensembling.

\textbf{Results:} 
RC-aware nnU-Netv2 ranked second in the BraTS 2025 Lighthouse Challenge Task 1 (BraTS-GLI 2025), with segmentation performance comparable to that of the first-place method, which relied on on-the-fly GliGAN-based synthetic tumor augmentation and multi-model ensembling. 
On the official  blind test set, our method  achieved   mean lesion-wise Dice scores of 0.7878, 0.8709, 0.7923, and 0.8715 and mean NSD@1.0 scores of 0.8309, 0.8712, 0.7980, and 0.8336 for enhancing tumor, RC, tumor core, and whole tumor, respectively. In paired post-treatment analysis, the RC-aware joint model improved mean lesion-wise Dice by 0.042 (95\% CI: [0.028, 0.057]) and mean NSD@1.0 by 0.043 (95\% CI: [0.028, 0.059]) compared with the joint baseline.

 \textbf{Conclusions:} RC-aware nnU-Netv2 provides a simple yet effective approach to pre- and post-treatment glioma segmentation, achieving competitive performance without synthetic tumor augmentation or multi-model ensembling. The results further demonstrate the benefits of explicitly handling empty region targets in joint learning from pre- and post-treatment MRI data.

}


\maketitle

% =========================================================
\section*{Introduction}

Gliomas are the most common malignant primary tumors of the adult central nervous system and show substantial heterogeneity in grade, molecular profile, spatial extent, and treatment response. 
Magnetic resonance imaging (MRI) is central to their diagnosis, surgical and radiotherapy planning, longitudinal monitoring, and response assessment. 
\textcolor{black}{ In routine practice and clinical trials, clinicians often need to distinguish enhancing tumor, non-enhancing tumor
tissue, surrounding abnormal signal, and treatment-related structures such as the
resection cavity (RC).}
Manual delineation of these regions is labor intensive and subject to inter-reader variability, whereas automated segmentation can provide reproducible volumetric measurements for treatment planning, follow-up evaluation, and quantitative imaging research \citep{menze2015brats,bakas2017tcga,bakas2018brats,rudie2022longitudinal,baig2024prepost}.

The Brain Tumor Segmentation (BraTS) Challenges have played a central role in 
advancing and standardizing automated glioma segmentation from MRI~\citep{menze2015brats,bakas2018brats}. 
Early BraTS tasks established multimodal MRI segmentation of preoperative tumor subregions as a common benchmark, using 
T1-weighted (T1), post-contrast
T1-weighted (T1Gd), T2-weighted (T2), and T2 Fluid-Attenuated Inversion
Recovery (T2-FLAIR) images~\citep{menze2015brats}. 
Subsequent BraTS releases expanded the scale, diversity, and annotation resources available to the community, including expert labels for The Cancer Genome Atlas (TCGA) glioma MRI collections and large multi-institutional preoperative cohorts \citep{bakas2017tcga,bakas2018brats,baid2021brats}. 
These efforts helped define clinically meaningful region groupings such as enhancing tumor, tumor core, and whole tumor, and established a rigorous framework for comparing automated segmentation methods.

\textcolor{black}{BraTS 2025 Lighthouse Challenge Task 1 (BraTS-GLI 2025) extends this setting by evaluating automated glioma segmentation on both pre-treatment and post-treatment MRI~\citep{brats_lighthouse_2025,deverdier2024brats,baig2024prepost,bakas2026permranker}.}
Pre-treatment segmentation focuses on the tumor and surrounding abnormal signal before surgical intervention, where the output can support baseline assessment and treatment planning. 
Post-treatment segmentation is more complex: surgery, radiation, systemic therapy, edema evolution, blood products, and postoperative anatomy alter image appearance, and the RC becomes a clinically meaningful structure that must be distinguished from residual or recurrent tumor. 
Recent studies on post-treatment glioma segmentation and combined pre-treatment and post-treatment segmentation have emphasized that automated delineation of RCs and longitudinal tumor subregions can reduce measurement burden and improve consistency in response assessment~\citep{rudie2022longitudinal,baig2024prepost,deverdier2024brats}.

This combined pre-treatment and post-treatment setting introduces a specific modeling difficulty. 
The RC is a valid target in post-treatment cases but is not anatomically applicable in pre-treatment scans. 
Thus, BraTS-GLI 2025~\citep{brats_lighthouse_2025} contains an imaging distribution shift between the two cohorts, and the target definitions also differ: ET, TC, and WT are applicable to both cohorts, whereas RC is applicable only to post-treatment cases.
A naive model trained on pooled pre-treatment and post-treatment data may learn unstable RC behavior, including false RC activation in pre-treatment cases or reduced RC sensitivity in post-treatment cases. 
At the same time, the official evaluation emphasizes lesion-wise Dice and lesion-wise normalized surface Dice (NSD) at a 1.0\,mm tolerance (NSD@1.0), placing greater weight on lesion detection and boundary accuracy than conventional voxel-wise overlap alone. 
These properties motivate an objective that handles empty RC targets
during pooled training, together with treatment-status-guided routing at
inference.

Methods submitted to
the Challenge  often improve leaderboard performance through multi-model ensembling, multi-fold fusion, extensive augmentation, or architecture combinations. 
The first-place BraTS-GLI 2025 method used on-the-fly GliGAN-based synthetic tumor  augmentation and submitted an ensemble of multiple nnU-Net-based models \citep{jain2026onthefly}. 
Two methods were ranked third: one used a customized EGASegNet architecture \citep{jin2026egasegnet}, and the other used the Medical Imaging Segmentation Toolkit (MIST) with systematic post-processing~\citep{celaya2025mist}.
\textcolor{black}{  Although these designs are effective, they may increase inference
complexity through ensembling, customized architectures,
challenge-specific augmentation, or post-processing pipelines.} 

\begin{figure}[b!]
\centering
\includegraphics[width=\textwidth]{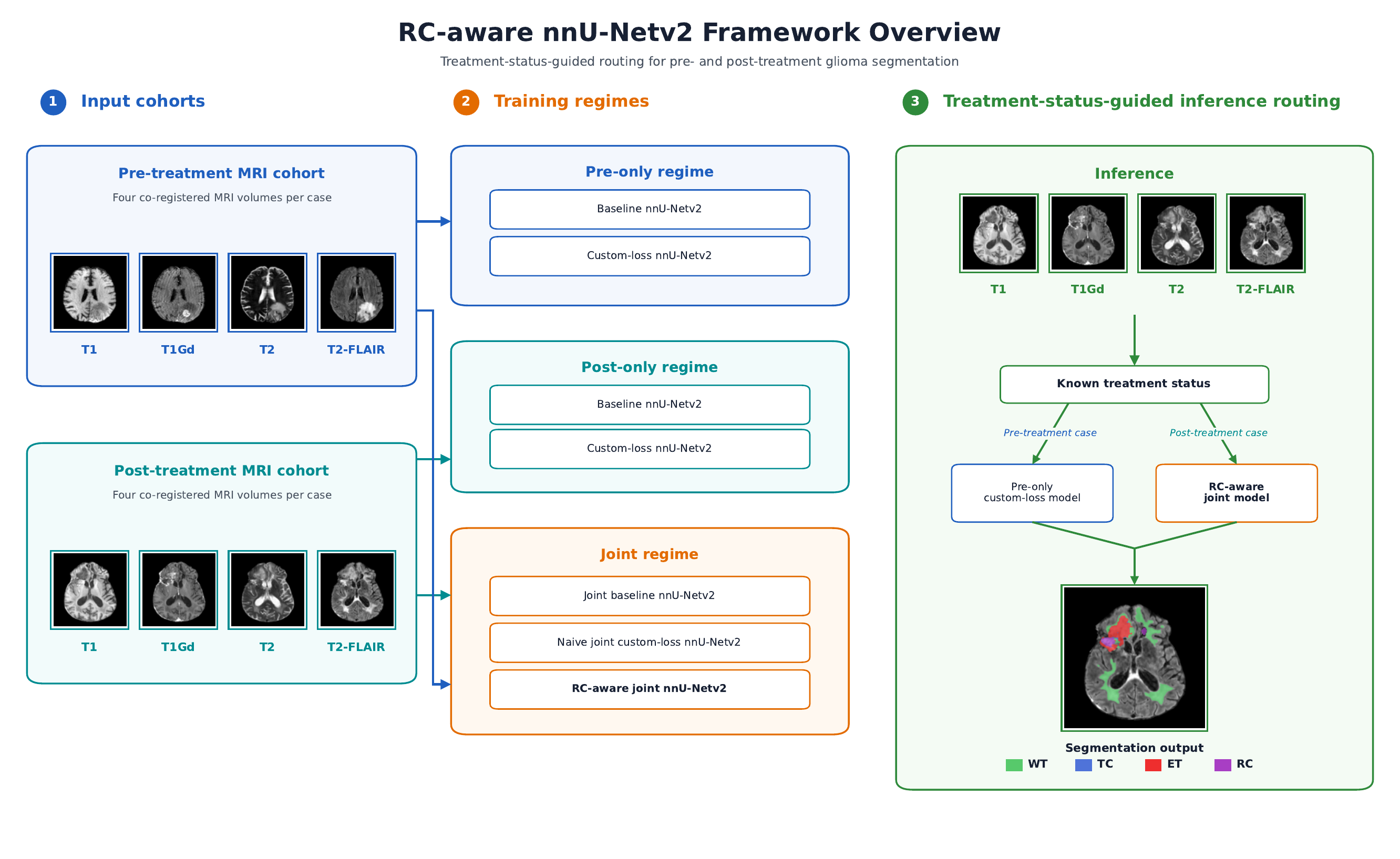}
\caption{Overview of the proposed RC-aware nnU-Netv2 framework. Three
training regimes are evaluated: pre-only, post-only, and joint training
on pooled data. During joint training, RC-aware supervision penalizes spurious RC activation in pre-treatment samples while retaining standard
RC supervision for post-treatment samples. At inference, the known
treatment status routes each case to a single model: pre-treatment cases
to the pre-only custom-loss model and post-treatment cases to the
RC-aware joint model.}
\label{fig:framework}
\end{figure}

\textcolor{black}{ 
In contrast, we propose RC-aware nnU-Netv2, 
a simple yet effective RC-aware framework based on nnU-Netv2~\citep{isensee2021nnunet,nnunetv2} 
for joint learning from 
pre- and post-treatment MRI data. Our submission, under the team name ECNU\_NYU, achieved second place in BraTS-GLI~2025, with segmentation performance comparable to that of the first-place method. Unlike the first-place method, which relied on challenge-specific synthetic tumor augmentation and multi-model ensembling, our approach uses only the standard nnU-Netv2 data augmentation  and a single specialized model per case.
\iffalse
pre-treatment and post-treatment MRI data. 
Our submission, under the team name ECNU\_NYU, 
achieved second place
in BraTS-GLI 2025, with segmentation performance comparable to that of the first-place method.
Our framework uses a joint
objective that handles empty region targets, including the structurally
empty RC target in pre-treatment cases, together with treatment-status-guided
case-level routing at inference. It achieves competitive performance with
the standard nnU-Netv2 augmentation 
and a single routed model per case, without challenge-specific synthetic
augmentation or multi-model ensembling used by the first-place method.}\fi
%Although these designs are effective, they may increase storage requirements, inference complexity, or dependence on challenge-specific augmentation.  In contrast, our method uses the pre-treatment and post-treatment distinction and the RC label asymmetry to obtain competitive performance without challenge-specific synthetic augmentation or multi-model fusion.
As illustrated in Figure~\ref{fig:framework}, 
RC-aware nnU-Netv2 incorporates two key training strategies. First, for single-cohort training, we introduce a lesion- and boundary-aware loss designed to better align model optimization with lesion-wise Dice and surface-based evaluation metrics. Second, for joint learning from pooled pre- and post-treatment data, we develop an RC-aware objective that applies standard Dice supervision to non-empty target channels and a controlled false-positive penalty to empty target channels, explicitly accommodating the structurally absent RC in pre-treatment cases. At inference, treatment-status-guided routing selects a specialized model for each case.
%\textcolor{black}{\st{ Because the official validation labels are not released to participating teams, we further construct a patient-level internal split over the union of pre-treatment and post-treatment cohorts, ensuring that all timepoints from the same patient remain in the same subset and preventing cross-timepoint leakage. This split is used for post-challenge ablation rather than as an exact reproduction of the official server-based model-selection process.}}

The main contributions of this work are summarized as follows:
\begin{itemize}[leftmargin=1em]
    \item 
We present RC-aware nnU-Netv2, a simple yet effective glioma segmentation framework that achieved second place in BraTS-GLI 2025 with performance comparable to that of the first-place method, using only the standard nnU-Netv2 data augmentation  and a single specialized model per case.

    \item We develop lesion- and boundary-aware single-cohort training and an RC-aware joint objective that accounts for empty region targets when learning from pooled pre-treatment and post-treatment data.

    \item We conduct a post-challenge evaluation incorporating patient-level data leakage prevention, controlled configuration-level ablations, and paired bootstrap analysis. The proposed RC-aware joint model demonstrates improved post-treatment segmentation performance compared with nnU-Netv2-based comparators.

    \item We compare 
    RC-aware nnU-Netv2 with strong competitors, 
    Swin UNETR~\citep{hatamizadeh2022swinunetr} and SegMamba~\citep{xing2024segmamba}, on 
     internal held-out test data, 
    demonstrating superior performance  
across most evaluated region--metric combinations while using fewer active model parameters per case.
\end{itemize}

% =========================================================
\section*{Related Work}

\subsection*{Strong medical image segmentation baselines}

U-Net~\citep{ronneberger2015unet} and its 3D variants are widely used architectures in medical image segmentation. 
Among them, nnU-Net~\citep{isensee2021nnunet} is especially influential because it standardizes preprocessing, architecture adaptation, deep supervision, and post-processing, thereby providing a strong and reproducible baseline across diverse biomedical datasets. 
{\color{black}
As a complete reimplementation of the original nnU-Net, 
 nnU-Netv2 \citep{nnunetv2} offers 
improved code structure and extensibility.}
The recent nnU-Net revisited study~\citep{isensee2024revisited} further underscores the importance of careful validation against strong baselines in 3D segmentation research. 
For BraTS-style brain tumor segmentation, nnU-Net-based pipelines remain
highly competitive, as demonstrated by recent top-ranked challenge
solutions~\citep{ferreira2024we,jain2026onthefly}.

\subsection*{Pre-treatment and post-treatment glioma segmentation}

Post-treatment glioma segmentation is often more difficult than pre-treatment segmentation because MRI appearance is altered by surgery, RCs, edema evolution, radiation effects, and other treatment-related changes. 
In addition, the target definitions differ across treatment statuses: the RC is meaningful for post-treatment scans but is not anatomically applicable before surgery. \textcolor{black}{ Prior work highlights the need to explicitly account for surgically altered regions in post-treatment segmentation~\citep{baig2024prepost}, and the BraTS annotation protocol formalizes
the RC as a separate segmentation target~\citep{deverdier2024brats}.}
%Prior work has emphasized that post-treatment segmentation requires explicit treatment of surgically altered regions and includes the RC as a meaningful target \citep{baig2024prepost}. 
Our work addresses this target-definition difference through a shared
four-region output space and a joint objective that explicitly handles
empty target channels during pooled training.

\subsection*{Boundary-aware and lesion-aware optimization}

Dice-based losses are widely used to address class imbalance in medical image segmentation \citep{milletari2016vnet}, and boundary-aware losses have been proposed to improve surface agreement in highly unbalanced segmentation tasks \citep{kervadec2021boundary}. 
These prior losses motivate objectives that emphasize foreground structure and boundary precision, which are relevant to BraTS-GLI  2025 because lesion-wise Dice and lesion-wise NSD@1.0 are primary evaluation metrics. 
However, the exact lesion-wise Dice used by the Challenge involves connected-component matching and is not straightforward to optimize directly during training. 
Our custom loss therefore uses differentiable surrogate terms: a Gaussian-smoothed target-weighted Dice component to emphasize lesion structure and a boundary-weighted binary cross-entropy  component to emphasize target boundaries.

\subsection*{Top-ranked BraTS-GLI 2025 solutions}

%The first-place method \citep{jain2026onthefly} used nnU-Net with on-the-fly synthetic tumor augmentation based on GliGAN~\citep{ferreira2024we}.  It trained a baseline model, a regular augmented model, and a customized augmented model, and selected the ensemble of all three models for final submission.  The reported test lesion-wise Dice scores were 0.804, 0.889, 0.800, and 0.877 for enhancing tumor (ET), resection cavity (RC), tumor core (TC), and whole tumor (WT), respectively. The corresponding NSD@1.0 scores were 0.843, 0.889, 0.805, and 0.842.

%One third-place method, \emph{EGASegNet} \citep{jin2026egasegnet}, proposed a customized 3D encoder--decoder network with an adaptive edge-guided module, multi-scale directional attention fusion, and an extreme group-aware attention former. 
%On the official test set, the reported mean lesion-wise Dice scores were 0.8009, 0.8262, 0.7988, and 0.8649 for ET, RC, TC, and WT, respectively. 
%The corresponding mean NSD@1.0 scores were 0.8444, 0.8285, 0.8078, and 0.8311.

%The other third-place method used MIST, a configurable medical image segmentation toolkit, and emphasized reproducible training and modular post-processing strategies such as small-component removal, label replacement, largest-component selection, hole filling, and morphological closing \citep{celaya2025mist}. 
%Its reported test lesion-wise Dice scores were 0.7875, 0.8400, 0.7873, and 0.8577 for ET, RC, TC, and WT, respectively.
%The corresponding NSD@1.0 scores were 0.8341, 0.8353, 0.7947, and 0.8210.

\textcolor{black}{The first-place method \citep{jain2026onthefly} combined nnU-Net with
on-the-fly synthetic tumor augmentation based on GliGAN
\citep{ferreira2024we}. It trained a baseline model, a regularly
augmented model, and a customized augmented model, and submitted the
ensemble of all three. One of the two third-place methods,
{EGASegNet} \citep{jin2026egasegnet}, proposed a customized 3D
encoder--decoder network with an adaptive edge-guided module,
multi-scale directional attention fusion, and an extreme group-aware
attention former. The other used MIST, a configurable medical image
segmentation toolkit, and emphasized reproducible training together with
modular post-processing such as small-component removal, label
replacement, largest-component selection, hole filling, and
morphological closing \citep{celaya2025mist}. Quantitative results for
all four top-ranked submissions are compared in
Table~\ref{tab:top3_full}.}

\textcolor{black}{ These methods differ from ours in emphasis: they improve
performance through architectural design, augmentation, ensembling, or
post-processing. Our submission instead focuses on the treatment-status
structure of the task, using an empty-target-aware joint objective during training
and case-level routing at inference, so that exactly one network
processes each case.}

%These methods differ from our submission, which focuses on timepoint-aware supervision and case-level routing rather than a new architecture, extensive augmentation, or model fusion. At inference, each case is assigned to a single branch according to its treatment status: pre-treatment cases are processed by the pre-treatment branch, and post-treatment cases are processed by the post-treatment RC-aware branch.
% =========================================================
\section*{Methods}\label{sec: method}

\subsection*{Problem setup and region representation}

Each case provides four co-registered MRI volumes
(T1, T1Gd, T2, and T2-FLAIR), which are
 stacked channel-wise
into a full-volume input tensor
$
\mathbf{x}_{\mathrm{full}}\in\mathbb{R}^{4\times H_0\times W_0\times D_0}.
$
The released BraTS-GLI 2025 volumes are distributed after co-registration to a common anatomical template, resampling to 1~mm isotropic resolution, and skull-stripping~\citep{menze2015brats,bakas2018brats,deverdier2024brats}. 
For the released full-volume images, the spatial matrix is $H_0\times W_0\times D_0=240\times240\times155$. 
During nnU-Netv2 training, patches are sampled from the preprocessed full volumes according to the nnU-Netv2 planning configuration; therefore, the network input patch size differs from the full-volume  size.
%The BraTS-GLI annotation uses mutually exclusive voxel labels for non-enhancing tumor core/necrosis (NETC/NCR), surrounding non-enhancing FLAIR hyperintensity (SNFH), enhancing tumor (ET), and resection cavity (RC). 
The BraTS-GLI annotation assigns each voxel to exactly one of five
mutually exclusive classes: background (label 0), non-enhancing tumor
core (NETC; label 1), surrounding non-enhancing T2-FLAIR hyperintensity
(SNFH; label 2), enhancing tissue (ET; label 3), and RC (label 4) \citep{deverdier2024brats,baig2024prepost}. The RC label is annotated only in post-treatment
cases. %In earlier pre-treatment BraTS releases, label 1 was termed the necrotic tumor core (NCR) and label 2 the peritumoral edematous/infiltrated tissue (ED); we use the unified BraTS-GLI nomenclature throughout this paper.

For region-based network training, these mutually exclusive labels are
converted into four overlapping targets
\citep{deverdier2024brats,baig2024prepost}. ET corresponds to label 3;
TC contains labels 1 and 3; WT contains labels 1, 2, and 3; and RC
corresponds to label 4. The tumor regions satisfy
$\mathrm{ET}\subseteq\mathrm{TC}\subseteq\mathrm{WT}$, whereas RC is
a separate region and is not included in TC or WT. The full official
BraTS-GLI evaluation includes ET, NETC, SNFH, RC, TC, and WT. The
region-level tables in this paper report the selected subset ET, RC,
TC, and WT. Label 3 is annotated as enhancing tissue, whereas the
corresponding evaluation region is conventionally termed enhancing
tumor; both are abbreviated ET.

Following the region-based training scheme of nnU-Netv2
\citep{isensee2021nnunet}, the network predicts four sigmoid-activated
channels in the fixed implementation order WT, TC, ET, and RC. At
inference, each channel is thresholded at 0.5. A categorical map is
constructed by assigning the thresholded masks sequentially in the
order WT, TC, ET, and RC, such that later assignments overwrite earlier
ones. RC therefore takes precedence over overlapping tumor-region
predictions. The remaining WT-only voxels are mapped to SNFH (official
label 2), the remaining TC-only voxels to NETC (official label 1), ET
to label 3, and RC to label 4. The submission pipeline removes
26-connected WT and RC components smaller than 50 voxels; no
component-size removal is applied to TC or ET.

Let
$
\mathbf{y}_{\mathrm{full}}\in
\{0,1\}^{4\times H_0\times W_0\times D_0}
$
denote the full-volume four-channel training target. During patch-based
training, paired image and target patches are sampled from
$\mathbf{x}_{\mathrm{full}}$ and $\mathbf{y}_{\mathrm{full}}$; below,
$H\times W\times D$ and $N$ denote the spatial dimensions and voxel
count of one patch. Because the training regions overlap, one voxel may
be positive in multiple channels.

For post-treatment cases, all four training targets are available. For
pre-treatment cases, RC is not anatomically applicable and its target
channel is stored as an all-zero map. This structurally empty RC target
motivates the empty-target handling in Eqs.~(\ref{eq:joint_loss})
and~(\ref{eq:vdice_loss}).

\subsection*{Baseline nnU-Netv2 objective}
The baseline nnU-Netv2 trainer uses the standard region-based objective
with deep supervision across multiple output resolutions
\cite{isensee2021nnunet}. It combines binary cross-entropy with logits (BCE-with-logits) and a
soft Dice term \cite{milletari2016vnet}, applied independently to each
region channel and weighted equally:
\begin{equation}
\mathcal{L}_{\mathrm{base}}
=
\mathcal{L}_{\mathrm{BCE}}
+
\mathcal{L}_{\mathrm{Dice}}.
\label{eq:base_loss}
\end{equation}
Let $p_i^{(c)}=\sigma(z_i^{(c)})$ denote the predicted probability
obtained from logit $z_i^{(c)}$, and let $y_i^{(c)}\in\{0,1\}$ denote
the target at voxel $i$ for region channel $c$. Because the region
targets overlap, each channel is activated by an independent sigmoid
$\sigma(\cdot)$ rather than by a softmax across channels. With
$C=4$ region channels and $N$ voxels, the two terms are defined as
\begin{equation}
\mathcal{L}_{\mathrm{BCE}}
=
-\frac{1}{CN}\sum_{c=0}^{C-1}\sum_{i=1}^{N}
\Bigl[
y_i^{(c)}\log p_i^{(c)}
+\bigl(1-y_i^{(c)}\bigr)\log\bigl(1-p_i^{(c)}\bigr)
\Bigr],
\label{eq:bce_loss}
\end{equation}
\begin{equation}
\mathcal{L}_{\mathrm{Dice}}
=
1-\frac{1}{C}\sum_{c=0}^{C-1}
\frac{
2\sum_{i=1}^{N}p_i^{(c)}y_i^{(c)}+\epsilon
}{
\sum_{i=1}^{N}p_i^{(c)}
+
\sum_{i=1}^{N}y_i^{(c)}
+
\epsilon
},
\label{eq:soft_dice_loss}
\end{equation}
where $\epsilon=10^{-5}$ is used throughout for numerical stability. Equations~(\ref{eq:bce_loss}) and~(\ref{eq:soft_dice_loss}) are written
for one sample. In the baseline implementation, the resulting
per-sample losses are averaged over the mini-batch.
The sum in Eq.~(\ref{eq:soft_dice_loss}) runs over all $C$ channels,
including channels whose target is empty; the joint objective introduced
in Eq.~(\ref{eq:joint_loss}) instead restricts the Dice term to
non-empty channels. This distinction is relevant for the RC channel of
pre-treatment cases.

Under deep supervision, the objective is evaluated at the supervised
decoder outputs and combined as
$\sum_{s}w_{s}\mathcal{L}_{\mathrm{base}}^{(s)}$. The weights are
initialized to halve at each lower resolution, the coarsest output is
assigned zero weight, and the remaining weights are normalized to sum
to one, following the nnU-Netv2 implementation \cite{isensee2021nnunet}. The
same deep-supervision scheme is used for all objectives in this work.
We refer to this configuration as \textbf{Baseline-nnU-Netv2}; it serves
as the reference method throughout the study, and no RC-specific
handling is applied to pre-treatment cases.

All nnU-Netv2-based models compared in this work share the same
preprocessing, data augmentation, network configuration, and
deep-supervision structure unless otherwise specified. They differ in
their complete training objectives. The comparisons therefore control
for architecture and data-pipeline factors but do not isolate the
contribution of any single loss component.

%The baseline nnU-Netv2 trainer uses the standard region-based objective with deep supervision across multiple output resolutions \citep{isensee2021nnunet}.  This objective combines Dice-style overlap \citep{milletari2016vnet} and binary cross-entropy with logits (BCE-with-logits) across region channels.  We refer to this configuration as \textbf{Baseline-nnU-Netv2}, which serves as the reference method throughout our study. 
%All custom models keep the same nnU-Netv2 preprocessing, data augmentation, architecture configuration, and deep supervision structure unless otherwise specified, so that the effect of the loss and supervision design can be isolated.

\subsection*{Single-cohort lesion- and boundary-aware loss}

For pre-only and post-only training, we use a composite objective that
combines the region-wise binary cross-entropy
$\mathcal{L}_{\mathrm{BCE}}$ of Eq.~(\ref{eq:bce_loss}), a lesion-aware
Dice surrogate $\mathcal{L}_{\mathrm{les}}$, and a boundary-weighted
binary cross-entropy term $\mathcal{L}_{\mathrm{bd}}$, both defined
below. The exact combination and weighting coefficients are specific to
this work, whereas the Dice-style overlap and boundary-weighting
components are motivated by prior Dice-based and boundary-aware
segmentation losses \cite{milletari2016vnet,kervadec2021boundary}. The
loss is defined as
\begin{equation}
\mathcal{L}_{\mathrm{custom}}
=
0.5\mathcal{L}_{\mathrm{BCE}}
+
0.5\mathcal{L}_{\mathrm{les}}
+
0.3\mathcal{L}_{\mathrm{bd}}.
\label{eq:custom_loss}
\end{equation}
The coefficients are empirical loss weights and are not constrained to
sum to one; the resulting loss scale was used consistently in all
single-cohort custom-loss experiments.

The lesion-aware term combines a voxel-wise Dice score with a
Gaussian-smoothed target-weighted Dice score. In the custom objectives,
the per-channel voxel-wise Dice is aggregated over both the mini-batch
and spatial dimensions:
\begin{equation}
D_{\mathrm{pixel}}^{(c)}
=
\frac{
2\sum_{b=1}^{B}\sum_{i=1}^{N}p_{b,i}^{(c)}y_{b,i}^{(c)}+\epsilon
}{
\sum_{b=1}^{B}\sum_{i=1}^{N}p_{b,i}^{(c)}
+
\sum_{b=1}^{B}\sum_{i=1}^{N}y_{b,i}^{(c)}
+\epsilon
},
\label{eq:custom_pixel_dice}
\end{equation}
where $B$ is the mini-batch size. The same definition is used for the
non-empty channels in Eq.~(\ref{eq:vdice_loss}). To obtain a
differentiable lesion-structure surrogate, the target mask is convolved
with a normalized $5\times5\times5$ Gaussian kernel $G_{\sigma}$ with
$\sigma=1.5$ voxels, and the resulting foreground weight map is
\begin{equation}
w_i^{(c)}
=
y_i^{(c)}
\frac{
\left(G_{\sigma} * y^{(c)}\right)_i
}{
\max_j \left(G_{\sigma} * y^{(c)}\right)_j+\epsilon
}.
\label{eq:lesion_weight}
\end{equation}
The Gaussian weight map is constructed independently for each sample.
For loss aggregation, the batch and spatial dimensions are subsequently
flattened, with $M=BN$ denoting the number of elements in one channel
across the mini-batch.
Because $w_i^{(c)}$ is normalized by the maximum of the smoothed mask,
its values lie between zero and one on the target foreground. The map
assigns larger weights to voxels with stronger local foreground support
and smaller weights to voxels near thin structures or lesion boundaries. 
This term is used as a differentiable lesion-structure surrogate rather
than as a direct implementation of the connected-component matching used
by the official lesion-wise Dice. The weighted Dice score is
\begin{equation}
D_{\mathrm{weighted}}^{(c)}
=
\frac{
2\sum_{i=1}^{M}\left(w_i^{(c)}\right)^{2}p_i^{(c)}y_i^{(c)}+\epsilon
}{
\sum_{i=1}^{M}w_i^{(c)}p_i^{(c)}
+
\sum_{i=1}^{M}w_i^{(c)}y_i^{(c)}
+
\epsilon
}.
\label{eq:weighted_dice}
\end{equation}
Because the target-derived weights are zero outside the foreground,
$D_{\mathrm{weighted}}^{(c)}$ primarily measures weighted foreground
recovery and does not directly penalize extra-target false positives;
these are penalized by $D_{\mathrm{pixel}}^{(c)}$ and BCE.
The squared weight in the numerator and the linear weight in the
denominator follow the implementation used in all experiments; this
asymmetry increases the sensitivity of $D_{\mathrm{weighted}}^{(c)}$ to
correctly predicted high-weight voxels. The lesion-aware Dice score and
the corresponding loss are
\begin{equation}
D_{\mathrm{les}}^{(c)}
=
0.7D_{\mathrm{weighted}}^{(c)}
+
0.3D_{\mathrm{pixel}}^{(c)},
\qquad
\mathcal{L}_{\mathrm{les}}
=
1-\frac{1}{C}\sum_{c=0}^{C-1}D_{\mathrm{les}}^{(c)}.
\label{eq:lesion_loss}
\end{equation}

The boundary term is a boundary-weighted BCE loss. For each sample and channel, $b_i^{(c)}\in\{0,1\}$ is computed from
absolute first-order target-mask differences along the three spatial
axes. The three difference maps are zero-padded on their leading sides,
summed, and clipped to $[0,1]$ to preserve the original tensor shape. The boundary loss is
\begin{equation}
\mathcal{L}_{\mathrm{bd}}
=
-\frac{1}{CM}
\sum_{c=0}^{C-1}\sum_{i=1}^{M}
\left(1+\lambda_{\mathrm{bd}}b_i^{(c)}\right)
\left[
y_i^{(c)}\log p_i^{(c)}
+
\left(1-y_i^{(c)}\right)\log\left(1-p_i^{(c)}\right)
\right],
\label{eq:boundary_loss}
\end{equation}
where $\lambda_{\mathrm{bd}}=2.0$ is the fixed boundary-weight
coefficient used in all experiments. Within $\mathcal{L}_{\mathrm{bd}}$, boundary voxels therefore receive three times the BCE weight assigned to
non-boundary voxels, which is intended to encourage better surface
agreement under NSD@1.0.

All coefficients in Eqs.~(\ref{eq:custom_loss}), (\ref{eq:lesion_loss})
and (\ref{eq:boundary_loss}), together with the Gaussian kernel
parameters, were fixed during method development and were not tuned
further for the post-challenge ablation; no sensitivity analysis was
performed.
Overall, this objective provides a differentiable surrogate for
lesion- and boundary-sensitive evaluation while preserving stable
end-to-end optimization.

\subsection*{RC-aware joint supervision}

For pooled pre-treatment and post-treatment training, all samples share
four output channels corresponding to WT, TC, ET, and RC. The RC target
is identically zero for pre-treatment samples, whereas post-treatment
samples may contain an annotated RC. We therefore use a joint objective
that combines BCE-with-logits, voxel-wise sigmoid Dice, and
boundary-weighted BCE:
\begin{equation}
\mathcal{L}_{\mathrm{joint}}
=
0.3\mathcal{L}_{\mathrm{BCE}}
+
0.5\mathcal{L}_{\mathrm{vDice}}
+
0.2\mathcal{L}_{\mathrm{bd}}.
\label{eq:joint_loss}
\end{equation}

The Dice component applies the standard per-channel Dice loss to
non-empty target channels and a controlled false-positive penalty to
empty target channels:
\begin{equation}
\mathcal{L}_{\mathrm{vDice}}
=
\frac{1}{|\mathcal{C}_{+}|+|\mathcal{C}_{0}|}
\left[
\sum_{c\in\mathcal{C}_{+}}
\left(1-D_{\mathrm{pixel}}^{(c)}\right)
+
\lambda_{\mathrm{fp}}
\sum_{c\in\mathcal{C}_{0}}
\frac{1}{M}\sum_{j=1}^{M}p_j^{(c)}
\right],
\label{eq:vdice_loss}
\end{equation}
where $\mathcal{C}_{+}$ and $\mathcal{C}_{0}$ denote channels with
non-empty and empty targets, respectively, within the current
mini-batch, $M$ is the number of prediction elements in one channel
across the mini-batch, and $\lambda_{\mathrm{fp}}=0.1$.

For pre-treatment samples, the all-zero RC target directly penalizes
spurious RC activation through the BCE and boundary-weighted BCE terms.
When the RC target is empty throughout a mini-batch, the Dice component
additionally applies the controlled false-positive penalty above. When
the RC target is non-empty, the standard Dice term is used. Although
this empty-target rule is applied to all region channels, it is
particularly relevant to RC because RC is structurally absent from all
pre-treatment cases.

No treatment-status indicator is supplied to the joint network during
training; treatment status is used only for case-level routing at
inference. Compared with the single-cohort custom loss in
Eq.~(\ref{eq:custom_loss}), the joint objective uses a different Dice
formulation and different component weights. The ablation therefore
compares complete training configurations rather than isolating the
effect of an individual loss term. The RC-aware nnU-Netv2 framework is
illustrated in Figure~\ref{fig:framework}.

\subsection*{Training variants and inference strategy}
\label{sec:variants}
We study three experimental regimes:
\begin{enumerate}
    \item \textbf{Pre-only training}: train on the pre-treatment cohort and evaluate on the pre-treatment internal held-out test set;
    \item \textbf{Post-only training}: train on the post-treatment cohort and evaluate on the post-treatment internal held-out test set;
    \item \textbf{Joint training}: train on the union of pre-treatment and post-treatment cohorts and evaluate separately on the pre-treatment and post-treatment internal held-out test sets.
\end{enumerate}
\textcolor{black}{ Across these regimes we evaluate the following configurations:
Baseline-nnU-Netv2, trained separately on each cohort (pre-only,
post-only) and on the pooled cohort (denoted Joint Baseline); the
single-cohort custom-loss model of Eq.~(\ref{eq:custom_loss}) for
pre-only and post-only training; naive joint training, which applies the
same custom loss to pooled data without RC-specific handling; and the
RC-aware joint model of Eq.~(\ref{eq:joint_loss}).}

%Across these regimes, we evaluate four training configurations where applicable: Baseline-nnU-Netv2; the single-cohort custom-loss model using Eq.~(\ref{eq:custom_loss}) for pre-only and post-only training; naive joint training, which applies the same custom loss in Eq.~(\ref{eq:custom_loss}) to pooled pre-treatment and post-treatment data without RC-specific handling; and the RC-aware joint model using Eq.~(\ref{eq:joint_loss}).

The proposed \textbf{RC-aware nnU-Netv2} denotes the full framework used for the final BraTS-GLI 2025 submission, including the pre-treatment custom-loss branch, the post-treatment RC-aware joint branch, and case-level routing by treatment status.
The pre-treatment branch is trained with the custom loss in Eq.~(\ref{eq:custom_loss}).
The post-treatment branch is the RC-aware joint model trained with
Eq.~(\ref{eq:joint_loss}), including the empty-target handling defined
in Eq.~(\ref{eq:vdice_loss}) during pooled training.
Thus, the final submission combines two specialized predictors rather
than a single undifferentiated model.
\textcolor{black}{This routing strategy is not a model ensemble: each case is assigned to
exactly one network, with no probability averaging across models,
voting, multi-fold fusion, or multi-architecture fusion. Sliding-window
inference and test-time mirroring follow the nnU-Netv2 defaults and are
applied identically to all compared models.}

This routed configuration was selected during the Challenge based on results from the official validation leaderboard. 
In the available pre-treatment validation feedback, the pre-only custom-loss branch improved the mean lesion-wise Dice scores for ET and TC from 0.801 and 0.832 to 0.817 and 0.846, respectively, relative to the standard pre-only nnU-Netv2 branch, while maintaining comparable WT performance (0.881 vs. 0.882). 
In the available post-treatment official validation leaderboard feedback, the RC-aware joint branch improved the mean lesion-wise Dice scores for ET, TC, and RC from 0.728, 0.718, and 0.651 to 0.744, 0.724, and 0.691, respectively, relative to the joint baseline, while WT remained comparable with a slight decrease from 0.881 to 0.877.
The post-challenge internal split reported below is therefore used to explain and stress-test the design choices, not as the sole basis for selecting the submitted model combination.

% =========================================================
% Continue with Part II: Experiments, Results, Discussion, Conclusion, Declarations, and References.
% =========================================================
% Part II of the manuscript.
% Paste this content after Part I and before \end{document} if you want a single Overleaf file.
% =========================================================

\section*{Experiments}

\subsection*{Datasets}

We use the labeled pre-treatment and post-treatment cohorts released for BraTS-GLI 2025~\citep{brats_lighthouse_2025}. 
The pre-treatment cohort contains 1251 cases from 1133 unique patients, and the post-treatment cohort contains 1621 cases from 731 unique patients. 
There are 34 patients shared across the two cohorts. 
The sample sizes and roles of the training, validation, held-out test, and official leaderboard datasets are summarized in Table~\ref{tab:split}.

\subsection*{Dataset splits and leakage prevention}

For the Challenge submission, models are trained using the labeled official training set, and model selection is guided by the official online validation leaderboard, whose labels are not released. 
The final rank is determined on the official blind test set.

For post-challenge analysis, we first split the labeled official training set at the patient level into an internal training pool and an internal held-out test set. 
Within the internal training pool, a validation subset was used for training monitoring and checkpoint selection; this subset did not consume additional patients or cases beyond the internal training counts reported in Table~\ref{tab:split}. 
The internal held-out test set was used only after training for ablation and model comparison.
All scans from the same patient are assigned to the same subset to prevent cross-timepoint leakage. \textcolor{black}{For the 34 patients contributing to
both cohorts, the assignment was made jointly across cohorts, so that a
patient's pre-treatment and post-treatment scans always fall in the same
subset; no patient appears in the training set of one cohort and the
held-out test set of the other.}

\begin{table}[!htbp]
\caption{Dataset and split summary. \textcolor{black}{ The internal validation subset is
drawn from  the internal training pool for training monitoring and
checkpoint selection and does not consume additional patients or cases;
the internal training and internal held-out test sets therefore account
for all patients and cases in the official training set.}}
\label{tab:split}
\centering
\small
\setlength{\tabcolsep}{4pt}
\renewcommand{\arraystretch}{1.22}
\begin{tabular*}{\textwidth}{@{\extracolsep{\fill}}
>{\raggedright\arraybackslash}p{3.4cm}
>{\raggedright\arraybackslash}p{2.6cm}
>{\raggedright\arraybackslash}p{2.6cm}
>{\raggedright\arraybackslash}p{4.0cm}
@{}}
\toprule
Dataset or split & Pre-treatment & Post-treatment & Use \\
\midrule
Official training set & 1133 patients, 1251 cases & 731 patients, 1621 cases & Challenge training and post-challenge re-splitting \\
Official validation leaderboard set & Labels not released & Labels not released & Challenge-phase model selection \\
Official blind test set & Labels not released & Labels not released & Final Challenge ranking \\
Internal training set & 893 patients, 977 cases & 594 patients, 1306 cases & Post-challenge model training \\
Internal validation subset & 196 cases & 262 cases & Within-training monitoring and checkpoint selection \\
Internal held-out test set & 240 patients, 274 cases & 137 patients, 315 cases & Ablation and model comparison \\
\bottomrule
\end{tabular*}
\end{table}

\subsection*{Implementation details}

All nnU-Netv2-based models are trained using the nnU-Netv2 implementation in~\citep{Brats2324}, while following the standard nnU-Netv2 preprocessing, planning, and augmentation pipeline.
We retain region-based outputs and deep supervision in all experiments.
Auxiliary decoder outputs at multiple resolutions are supervised during
training; the coarsest output is assigned zero weight, and the remaining
deep-supervision weights are normalized to sum to one.
Unless otherwise specified, the preprocessing plan, patch size, network configuration, optimizer schedule, learning rate decay, mixed precision, and standard data augmentation are determined by the nnU-Netv2 self-configuring pipeline or kept at its default settings.
We do not use the repository's synthetic-tumor augmentation or ensemble submission pipeline.
The differences across nnU-Netv2-based ablation models are implemented
through customized trainer classes that define their respective
training objectives.
For our local ablation study, each prediction is generated from a single nnU-Netv2 checkpoint trained within the internal training set, and no local ablation experiment uses additional synthetic tumor generation, multi-fold fusion, or multi-model ensembling.

For final submission, the single containerized inference pipeline routes each test case according to the treatment-status information encoded in the challenge case identifier. Pre-treatment cases are processed by the pre-only custom-loss model and post-treatment cases by the RC-aware joint model; all predictions are written to a common output directory.
The submitted inference pipeline uses no \textcolor{black}{ model-level} probability averaging, majority voting, multi-fold fusion, multi-architecture ensemble, or on-the-fly GliGAN-style synthetic tumor augmentation.

\subsection*{Evaluation metrics}

For the testing-phase comparison in Table~\ref{tab:top3_full}, we report
lesion-wise Dice and NSD@1.0 for the selected regions ET, RC, TC, and
WT. The full official BraTS-GLI evaluation additionally includes NETC
and SNFH.
The official BraTS ranking is based on a rank-based protocol that aggregates case-wise rankings across metrics and regions, so the final rank is not determined solely by the arithmetic mean of the displayed scores \citep{bakas2026permranker}.

For post-challenge local ablation, all metrics are recomputed using the official-style BraTS-GLI lesion-wise evaluation code. 
We report lesion-wise Dice, lesion-wise NSD@1.0, and lesion-wise 
95th-percentile Hausdorff Distance (HD95). 
Values are reported as mean $\pm$ standard deviation over held-out test cases.
For paired post-treatment analysis, each metric was first averaged with
equal weight over ET, RC, TC, and WT within each of the 315 held-out
cases. Dice and NSD@1.0 differences were defined as the RC-aware joint
model minus the comparator, whereas HD95 differences were defined as
the comparator minus the RC-aware joint model. Positive differences
therefore always indicate better performance of the RC-aware joint
model. Infinite HD95 values were replaced by 337~mm before averaging.

We used a paired case-level percentile bootstrap~\citep{efron1993bootstrap} with $B=200{,}000$
resamples. Each replicate
sampled 315 cases with replacement while preserving the model pairing.
The 95\% confidence interval was defined by the 2.5th and 97.5th
percentiles of the bootstrap mean differences. Let
$n_{\mathrm{opp}}$ denote the number of bootstrap mean differences less
than or equal to zero. The two-sided bootstrap $p$ value was
\begin{equation}
p=\min\left(1,\frac{2(1+n_{\mathrm{opp}})}{B+1}\right).
\label{eq:bootstrap_p}
\end{equation}
The same bootstrap distribution was used for confidence intervals and
$p$ values. Reported $p$ values are nominal and unadjusted for multiple
comparisons.

\subsection*{Experimental design}
 
The training configurations defined under \textit{Training variants
and inference strategy} are
compared on the internal held-out test sets; the configuration names
match the Method column of Tables~\ref{tab:local_lwdice}--\ref{tab:local_lwhd95}, and
the Eval column indicates which held-out set each row refers to.
Single-cohort models appear once, on the matched cohort; jointly trained
models appear twice, once per cohort.

The central ablation question is whether the proposed joint training
configuration provides an advantage over standard pooled training and
naive application of the single-cohort custom loss. These experiments
compare complete training configurations: the RC-aware joint model
differs from naive joint training in its Dice formulation and component
weights. Consequently, the results should not be interpreted as
isolating the effect of an individual loss term.

Because the final treatment-status-routed submission was chosen using official
validation feedback during the challenge phase, the internal split is
interpreted as a post-hoc diagnostic evaluation. In particular, a
component that was useful for the official submission need not be the
best-performing component on every internal split, especially for
pre-treatment cases.

% =========================================================
\section*{Results}

\subsection*{Official testing-phase performance}
Table~\ref{tab:top3_full} summarizes the reported top-ranked BraTS-GLI 2025 results. The first-place method used a three-model nnU-Net
ensemble with on-the-fly GliGAN-based synthetic tumor augmentation
\cite{jain2026onthefly,ferreira2024we}, whereas our submission ranked
second using treatment-status-guided routing without additional synthetic
augmentation or multi-model ensembling. The two methods tied for third
place used different strategies: EGASegNet used a customized 3D
encoder--decoder architecture with edge-guided and attention modules
\cite{jin2026egasegnet}, whereas MIST used a configurable medical image
segmentation toolkit with modular post-processing \cite{celaya2025mist}.
Results are reported as mean $\pm$ standard deviation (SD) over the
official Challenge test set. The test data and labels were withheld from
the participating teams, and each submitted model was evaluated by the
Challenge organizers through a Docker container.

Across the four displayed regions (ET, RC, TC, and WT), our method
achieved arithmetic mean lesion-wise Dice and NSD@1.0 scores of 0.8306
and 0.8334, respectively. These descriptive means were 0.0117 and
0.0113 lower than those of the first-place ensemble. Compared with the
two third-place methods, our largest advantage was observed for RC,
where our lesion-wise Dice was 0.8709, compared with 0.8400 for MIST
and 0.8262 for EGASegNet. These cross-method comparisons are
descriptive.
Because the winning method relied on an ensemble with on-the-fly GliGAN
augmentation, these results indicate that modeling the pre-treatment and
post-treatment structure and the RC label semantics can remain
competitive with ensemble-based methods, \textcolor{black}{ even though each case is
processed by a single model rather than an ensemble.}

\begin{table}[htbp]
\caption{Reported top-ranked BraTS-GLI 2025 blind test results. Values are reported as mean $\pm$ SD. The official ranking follows the BraTS rank-based evaluation protocol rather than an arithmetic mean over regions.}
\label{tab:top3_full}
\centering
\tiny
\setlength{\tabcolsep}{1.8pt}
\renewcommand{\arraystretch}{1.15}
\begin{tabular}{c l l cccc}
\toprule
Rank & Method & Metric & ET & RC & TC & WT \\
\midrule
1
& On-the-fly GliGAN + nnU-Net 
& LW Dice & $0.8040\pm0.2635$ & $0.8886\pm0.2532$ & $0.7998\pm0.2869$ & $0.8768\pm0.1775$ \\
& & NSD@1.0 & $0.8426\pm0.2563$ & $0.8895\pm0.2465$ & $0.8051\pm0.2763$ & $0.8416\pm0.1942$ \\
\addlinespace
2
& ECNU\_NYU (ours)
& LW Dice & $0.7878\pm0.2751$ & $0.8709\pm0.2794$ & $0.7923\pm0.2903$ & $0.8715\pm0.1801$ \\
& & NSD@1.0 & $0.8309\pm0.2651$ & $0.8712\pm0.2742$ & $0.7980\pm0.2794$ & $0.8336\pm0.1971$ \\
\addlinespace
3 
& EGASegNet 
& LW Dice & $0.8009\pm0.2622$ & $0.8262\pm0.3257$ & $0.7988\pm0.2853$ & $0.8649\pm0.1852$ \\
& & NSD@1.0 & $0.8444\pm0.2503$ & $0.8285\pm0.3196$ & $0.8078\pm0.2711$ & $0.8311\pm0.1983$ \\
\addlinespace
3
& MIST
& LW Dice & $0.7875\pm0.2707$ & $0.8400\pm0.3128$ & $0.7873\pm0.2912$ & $0.8577\pm0.1928$ \\
& & NSD@1.0 & $0.8341\pm0.2563$ & $0.8353\pm0.3106$ & $0.7947\pm0.2761$ & $0.8210\pm0.2048$ \\
\bottomrule
\end{tabular}
\end{table}

\subsection*{Ablation study}
\label{sec:ablation}

\textcolor{black}{ Tables~\ref{tab:local_lwdice}--\ref{tab:local_lwhd95} summarize the local
ablation results in terms of lesion-wise Dice, NSD@1.0, and HD95. In the pre-treatment evaluation, the RC-aware joint model matched or
slightly exceeded the pre-only baseline in mean lesion-wise Dice (0.878
vs.\ 0.877), whereas the pre-only custom-loss model did not outperform
the baseline on this internal split.
In the post-treatment evaluation, the RC-aware joint model obtained the
best value in every region and every metric across
Tables~\ref{tab:local_lwdice}--\ref{tab:local_lwhd95}; in mean
lesion-wise Dice, it improved the score from $0.706\pm0.202$ for the
joint baseline to $0.748\pm0.194$. In contrast, naive joint custom-loss
training performed worst on the post-treatment held-out test set,
indicating that simply applying the single-cohort custom loss to pooled
pre-treatment and post-treatment data was insufficient and that the
proposed complete joint training configuration was more effective.}

For lesion-wise NSD@1.0, Table~\ref{tab:local_lwnsd} reports results using the same evaluation code. 
The post-treatment results follow the same pattern as lesion-wise Dice:
the RC-aware joint model improved the mean score from
$0.730\pm0.199$ for the joint baseline to $0.773\pm0.193$.
This indicates that the gain is not limited to volumetric overlap but also appears in the surface-tolerance metric used by the official evaluation.
On the pre-treatment split, the custom-loss and RC-aware variants showed smaller and less uniform differences, reinforcing that the strongest evidence for the proposed joint training
configuration comes from post-treatment cases.

For lesion-wise HD95, Table~\ref{tab:local_lwhd95} reports boundary-distance performance, where lower values indicate better performance. 
In the post-treatment evaluation, the RC-aware joint model achieved the lowest mean lesion-wise HD95, reducing it from $53.30\pm66.80$ for the joint baseline to $45.67\pm63.69$. 
In the pre-treatment evaluation, the RC-aware joint model also obtained the lowest mean lesion-wise HD95 among the joint-training models.

\begin{table}[!htbp]
\caption{Post-challenge local ablation: lesion-wise Dice computed with the BraTS-GLI lesion-wise evaluator. Values are mean $\pm$ standard deviation. The Mean column is computed per case over ET, TC, and WT for pre-treatment evaluation and over ET, RC, TC, and WT for post-treatment evaluation. \textcolor{black}{ Bold indicates the highest value in each column within each evaluation split (Pre or Post).}}
\label{tab:local_lwdice}
\centering
\tiny
\setlength{\tabcolsep}{1.2pt}
\renewcommand{\arraystretch}{1.15}
\begin{tabular}{llcccccc}
\toprule
Setting & Method & Eval & ET & RC & TC & WT & Mean \\
\midrule
Pre-only & Baseline-nnU-Netv2 & Pre & $0.844\pm0.214$ & \NA & $\mathbf{0.894\pm0.161}$ & $\mathbf{0.893\pm0.168}$ & $0.877\pm0.137$ \\
Pre-only & Custom loss & Pre & $0.843\pm0.205$ & \NA & $0.883\pm0.180$ & $0.887\pm0.168$ & $0.871\pm0.142$ \\
\addlinespace
Post-only & Baseline-nnU-Netv2 & Post & $0.632\pm0.357$ & $0.699\pm0.313$ & $0.631\pm0.353$ & $0.794\pm0.217$ & $0.689\pm0.199$ \\
Post-only & Custom loss & Post & $0.682\pm0.349$ & $0.720\pm0.316$ & $0.686\pm0.341$ & $0.788\pm0.230$ & $0.719\pm0.188$ \\
\addlinespace
Joint & Joint Baseline & Pre & $0.840\pm0.203$ & \NA & $0.878\pm0.185$ & $0.874\pm0.179$ & $0.864\pm0.139$ \\
Joint & Joint Baseline & Post & $0.691\pm0.345$ & $0.675\pm0.336$ & $0.665\pm0.343$ & $0.794\pm0.222$ & $0.706\pm0.202$ \\
\addlinespace
Joint & Naive Joint Custom & Pre & $0.812\pm0.221$ & \NA & $0.849\pm0.214$ & $0.850\pm0.195$ & $0.837\pm0.159$ \\
Joint & Naive Joint Custom & Post & $0.572\pm0.374$ & $0.620\pm0.334$ & $0.551\pm0.367$ & $0.744\pm0.242$ & $0.622\pm0.198$ \\
\addlinespace
Joint & RC-aware Joint & Pre & $\mathbf{0.853\pm0.195}$ & \NA & $0.892\pm0.168$ & $0.889\pm0.163$ & $\mathbf{0.878\pm0.135}$ \\
Joint & RC-aware Joint & Post & $\mathbf{0.737\pm0.325}$ & $\mathbf{0.725\pm0.320}$ & $\mathbf{0.718\pm0.327}$ & $\mathbf{0.813\pm0.220}$ & $\mathbf{0.748\pm0.194}$ \\
\bottomrule
\end{tabular}
\end{table}
%\FloatBarrier

\begin{table}[!htbp]
\caption{Post-challenge local ablation: lesion-wise NSD@1.0 computed with the BraTS-GLI lesion-wise evaluator. Values are mean $\pm$ standard deviation. The Mean column is computed per case over the evaluated regions. Bold indicates the highest value in each column within each evaluation split (Pre or Post).}
\label{tab:local_lwnsd}
\centering
\tiny
\setlength{\tabcolsep}{1.2pt}
\renewcommand{\arraystretch}{1.15}
\begin{tabular}{llcccccc}
\toprule
Setting & Method & Eval & ET & RC & TC & WT & Mean \\
\midrule
Pre-only & Baseline-nnU-Netv2 & Pre & $0.868\pm0.210$ & \NA & $\mathbf{0.858\pm0.195}$ & $\mathbf{0.838\pm0.188}$ & $\mathbf{0.855\pm0.158}$ \\
Pre-only & Custom loss & Pre & $0.862\pm0.205$ & \NA & $0.841\pm0.211$ & $0.827\pm0.191$ & $0.843\pm0.162$ \\
\addlinespace
Post-only & Baseline-nnU-Netv2 & Post & $0.694\pm0.355$ & $0.706\pm0.290$ & $0.683\pm0.347$ & $0.780\pm0.208$ & $0.716\pm0.202$ \\
Post-only & Custom loss & Post & $0.734\pm0.346$ & $0.719\pm0.301$ & $0.733\pm0.334$ & $0.782\pm0.227$ & $0.742\pm0.192$ \\
\addlinespace
Joint & Joint Baseline & Pre & $0.865\pm0.202$ & \NA & $0.840\pm0.201$ & $0.806\pm0.197$ & $0.837\pm0.157$ \\
Joint & Joint Baseline & Post & $0.748\pm0.338$ & $0.677\pm0.315$ & $0.712\pm0.335$ & $0.784\pm0.215$ & $0.730\pm0.199$ \\
\addlinespace
Joint & Naive Joint Custom & Pre & $0.843\pm0.214$ & \NA & $0.803\pm0.221$ & $0.761\pm0.203$ & $0.802\pm0.170$ \\
Joint & Naive Joint Custom & Post & $0.636\pm0.377$ & $0.600\pm0.312$ & $0.605\pm0.366$ & $0.731\pm0.229$ & $0.643\pm0.201$ \\
\addlinespace
Joint & RC-aware Joint & Pre & $\mathbf{0.878\pm0.189}$ & \NA & $0.855\pm0.194$ & $0.827\pm0.186$ & $0.854\pm0.152$ \\
Joint & RC-aware Joint & Post & $\mathbf{0.787\pm0.320}$ & $\mathbf{0.737\pm0.299}$ & $\mathbf{0.762\pm0.319}$ & $\mathbf{0.807\pm0.218}$ & $\mathbf{0.773\pm0.193}$ \\
\bottomrule
\end{tabular}
\end{table}
%\FloatBarrier

\begin{table}[h!]
\caption{Post-challenge local ablation: lesion-wise HD95 computed with the BraTS-GLI lesion-wise evaluator. Lower is better. Values are mean $\pm$ standard deviation. The Mean column is computed per case over the evaluated regions. Bold indicates the lowest value in each column within each evaluation split.}
\label{tab:local_lwhd95}
\centering
\tiny
\setlength{\tabcolsep}{1.0pt}
\renewcommand{\arraystretch}{1.15}
\begin{tabular}{llcccccc}
\toprule
Setting & Method & Eval & ET & RC & TC & WT & Mean \\
\midrule
Pre-only & Baseline-nnU-Netv2 & Pre & $20.88\pm63.67$ & \NA & $14.40\pm44.76$ & $19.58\pm53.18$ & $18.29\pm39.26$ \\
Pre-only & Custom loss & Pre & $20.25\pm61.03$ & \NA & $15.60\pm46.67$ & $21.05\pm53.49$ & $18.96\pm40.06$ \\
\addlinespace
Post-only & Baseline-nnU-Netv2 & Post & $74.82\pm124.36$ & $42.76\pm95.79$ & $74.84\pm123.07$ & $37.41\pm69.81$ & $57.46\pm69.78$ \\
Post-only & Custom loss & Post & $65.51\pm120.03$ & $43.26\pm100.46$ & $61.83\pm115.80$ & $42.69\pm75.72$ & $53.32\pm64.41$ \\
\addlinespace
Joint & Joint Baseline & Pre & $17.66\pm58.37$ & \NA & $\mathbf{12.42\pm40.29}$ & $23.60\pm57.72$ & $17.89\pm36.60$ \\
Joint & Joint Baseline & Post & $59.51\pm116.80$ & $51.36\pm106.64$ & $63.68\pm117.71$ & $38.67\pm70.90$ & $53.30\pm66.80$ \\
\addlinespace
Joint & Naive Joint Custom & Pre & $20.40\pm64.05$ & \NA & $12.52\pm38.33$ & $29.31\pm63.36$ & $20.74\pm40.53$ \\
Joint & Naive Joint Custom & Post & $86.90\pm134.26$ & $65.88\pm109.57$ & $91.38\pm134.86$ & $49.70\pm77.69$ & $73.46\pm73.35$ \\
\addlinespace
Joint & RC-aware Joint & Pre & $\mathbf{15.56\pm52.60}$ & \NA & $13.01\pm41.53$ & $\mathbf{19.36\pm52.34}$ & $\mathbf{15.98\pm35.42}$ \\
Joint & RC-aware Joint & Post & $\mathbf{50.89\pm109.02}$ & $\mathbf{40.51\pm94.15}$ & $\mathbf{55.32\pm109.49}$ & $\mathbf{35.95\pm71.98}$ & $\mathbf{45.67\pm63.69}$ \\
\bottomrule
\end{tabular}
\end{table}
%\FloatBarrier

\subsection*{Paired post-treatment statistical analysis}
\textcolor{black}{To assess whether the post-treatment improvements were consistent at the
case level, we performed paired comparisons between the RC-aware joint
model and three comparators: the joint baseline, naive joint custom-loss
training, and the post-only custom-loss model
(Table~\ref{tab:paired_stats}). All nine comparisons favored the
RC-aware joint model, with confidence intervals excluding zero
throughout.}

\textcolor{black}{The largest margin was observed against naive joint
custom-loss training ($+0.127$ in mean lesion-wise Dice), which was the
least favorable evaluated configuration for post-treatment segmentation.
The RC-aware joint configuration also outperformed the joint baseline
($+0.042$) and post-only custom-loss configuration ($+0.029$). Because
these models use different complete loss formulations, these comparisons
support the complete RC-aware joint configuration but do not isolate the
effect of pooled training or any individual loss component.}

%To assess whether the post-treatment improvements were consistent at the case level, we performed paired comparisons between the RC-aware joint model and three relevant comparators: the joint baseline, naive joint custom-loss training, and the post-only custom-loss model.
%Table~\ref{tab:paired_stats} reports mean paired improvements with paired bootstrap 95\% confidence intervals and bootstrap $p$ values.
%For lesion-wise Dice and NSD@1.0, positive values indicate higher scores for the RC-aware joint model; for lesion-wise HD95, positive values indicate a reduction in HD95 relative to the comparator.
%The Dice and NSD@1.0 improvements were statistically significant for all three comparisons, with bootstrap $p$ values between $1.0\times10^{-5}$ and $3.0\times10^{-5}$.
%Using the paired bootstrap analysis, HD95 reductions were also significant relative to the joint baseline ($p=0.00663$), naive joint custom-loss training ($p=1.0\times10^{-5}$), and the post-only custom-loss model ($p=0.00375$).

\begin{table}[h!]
\caption{\textcolor{black}{ Paired post-treatment comparison of the RC-aware joint model
against three comparators on the internal held-out test set
($n=315$ cases). Differences are means of the case-level paired differences after
equal-weight averaging over ET, RC, TC, and WT;
percentile 95\% confidence intervals and two-sided nominal, unadjusted
$p$ values are obtained from a paired bootstrap with 200{,}000 case-level resamples. Positive values favor the
RC-aware joint model: higher lesion-wise Dice and NSD@1.0, and a larger
reduction in HD95.}}
\label{tab:paired_stats}
\centering
\small
\setlength{\tabcolsep}{6pt}
\renewcommand{\arraystretch}{1.2}
\begin{tabular}{llccc}
\toprule
Comparator & & $\Delta$ LW Dice & $\Delta$ NSD@1.0 & HD95 reduction (mm)\\
\midrule
Joint Baseline & Difference & $+0.042$ & $+0.043$ & $+7.63$\\
& 95\% CI & [0.028, 0.057] & [0.028, 0.059] & [2.08, 13.26]\\
& $p$ value & $1.0\times10^{-5}$ & $1.0\times10^{-5}$ & $0.00663$\\
\addlinespace
Naive Joint Custom & Difference & $+0.127$ & $+0.130$ & $+27.79$\\
& 95\% CI & [0.105, 0.148] & [0.108, 0.152] & [19.63, 35.95]\\
& $p$ value& $1.0\times10^{-5}$ & $1.0\times10^{-5}$ & $1.0\times10^{-5}$\\
\addlinespace
Post-only Custom & Difference & $+0.029$ & $+0.031$ & $+7.65$\\
& 95\% CI & [0.015, 0.043] & [0.017, 0.046] & [2.37, 12.96]\\
& $p$ value& $3.0\times10^{-5}$ & $3.0\times10^{-5}$ & $0.00375$\\
\bottomrule
\end{tabular}
\end{table}
%\FloatBarrier

\subsection*{Comparison with additional architectures}

Table~\ref{tab:architecture_comparison} compares SegMamba~\citep{xing2024segmamba} and Swin UNETR~\citep{hatamizadeh2022swinunetr} with the two routed models used in our final submission under the same BraTS-GLI lesion-wise evaluation protocol. 
SegMamba is evaluated using a public PyTorch implementation adapted to the BraTS-GLI four-modality input and region-output setting. 
Swin UNETR is evaluated using a MONAI-based implementation with \texttt{feature\_size=48} and an input patch size of $4\times128\times128\times128$.

Across both splits, the proposed routed models obtained the best values in
a large majority of regions and metrics: on the post-treatment split
they led in every region except TC, where SegMamba was marginally better,
and on the pre-treatment split they led in all Dice and HD95 regions.
For pre-treatment NSD@1.0, Swin UNETR was slightly higher for ET and TC,
whereas SegMamba was slightly higher for WT. Notably, the routed
branches use 30.8\,M parameters, roughly half of Swin UNETR (62.2\,M) and
SegMamba (67.4\,M; Table~\ref{tab:model_complexity}), so this accuracy is
achieved at substantially lower parameter cost.
%Compared with these alternative backbones, the proposed routed models remain competitive across the internal held-out test sets, while the alternative backbones are stronger only for selected individual regions and metrics.

Because these additional-architecture experiments are exploratory and are not the primary paired ablation comparisons, we report descriptive mean $\pm$ standard deviation values rather than formal paired significance tests.
%Patch-level model complexity is reported in Table~\ref{tab:model_complexity} to contextualize computational cost across representative backbones. 
These estimates should be interpreted as architecture-level comparisons, because case-level inference time also depends on sliding-window tiling, overlap, implementation, and test-time processing.

\begin{table}[!htbp]
\caption{Architecture-level comparison on the internal held-out test set
using the BraTS-GLI lesion-wise evaluator. Values are mean $\pm$ standard
deviation; bold marks the best value in each column within each split and
metric. The proposed framework is represented by the two routed models
used in the final submission, Pre-only Custom for pre-treatment cases and
RC-aware Joint for post-treatment cases. }
\label{tab:architecture_comparison}
\centering
\tiny
\setlength{\tabcolsep}{1.0pt}
\renewcommand{\arraystretch}{1.12}
\begin{tabular}{ll l cccc}
\toprule
Model & Split & Metric & ET & RC & TC & WT \\
\midrule
Swin UNETR & Post & Lesion-wise Dice & $0.650\pm0.354$ & $0.630\pm0.339$ & $0.654\pm0.347$ & $0.691\pm0.268$ \\
Swin UNETR & Post & Lesion-wise NSD@1.0 & $0.702\pm0.358$ & $0.633\pm0.321$ & $0.703\pm0.345$ & $0.683\pm0.261$ \\
Swin UNETR & Post & Lesion-wise HD95 & $76.68\pm122.80$ & $69.07\pm114.34$ & $74.02\pm119.84$ & $77.11\pm92.38$ \\
\addlinespace
SegMamba & Post & Lesion-wise Dice & $0.733\pm0.322$ & $0.709\pm0.318$ & $\mathbf{0.724\pm0.320}$ & $0.799\pm0.228$ \\
SegMamba & Post & Lesion-wise NSD@1.0 & $0.782\pm0.319$ & $0.720\pm0.294$ & $\mathbf{0.769\pm0.312}$ & $0.800\pm0.222$ \\
SegMamba & Post & Lesion-wise HD95 & $52.57\pm108.62$ & $43.30\pm96.99$ & $\mathbf{53.86\pm107.15}$ & $39.10\pm73.72$ \\
\addlinespace
RC-aware Joint (ours) & Post & Lesion-wise Dice & $\mathbf{0.737\pm0.325}$ & $\mathbf{0.725\pm0.320}$ & $0.718\pm0.327$ & $\mathbf{0.813\pm0.220}$ \\
RC-aware Joint (ours) & Post & Lesion-wise NSD@1.0 & $\mathbf{0.787\pm0.320}$ & $\mathbf{0.737\pm0.299}$ & $0.762\pm0.319$ & $\mathbf{0.807\pm0.218}$ \\
RC-aware Joint (ours) & Post & Lesion-wise HD95 & $\mathbf{50.89\pm109.02}$ & $\mathbf{40.51\pm94.15}$ & $55.32\pm109.49$ & $\mathbf{35.95\pm71.98}$ \\
\addlinespace
Swin UNETR & Pre & Lesion-wise Dice & $0.838\pm0.217$ & \NA & $0.878\pm0.186$ & $0.878\pm0.179$ \\
Swin UNETR & Pre & Lesion-wise NSD@1.0 & $\mathbf{0.863\pm0.211}$ & \NA & $\mathbf{0.842\pm0.206}$ & $0.819\pm0.200$ \\
Swin UNETR & Pre & Lesion-wise HD95 & $20.53\pm62.34$ & \NA & $15.86\pm46.57$ & $24.19\pm58.97$ \\
\addlinespace
SegMamba & Pre & Lesion-wise Dice & $0.831\pm0.222$ & \NA & $0.869\pm0.199$ & $0.882\pm0.182$ \\
SegMamba & Pre & Lesion-wise NSD@1.0 & $0.858\pm0.216$ & \NA & $0.834\pm0.216$ & $\mathbf{0.828\pm0.197}$ \\
SegMamba & Pre & Lesion-wise HD95 & $22.85\pm66.56$ & \NA & $16.96\pm49.91$ & $23.61\pm60.05$ \\
\addlinespace
Pre-only Custom (ours) & Pre & Lesion-wise Dice & $\mathbf{0.843\pm0.205}$ & \NA & $\mathbf{0.883\pm0.180}$ & $\mathbf{0.887\pm0.168}$ \\
Pre-only Custom (ours) & Pre & Lesion-wise NSD@1.0 & $0.862\pm0.205$ & \NA & $0.841\pm0.211$ & $0.827\pm0.191$ \\
Pre-only Custom (ours) & Pre & Lesion-wise HD95 & $\mathbf{20.25\pm61.03}$ & \NA & $\mathbf{15.60\pm46.67}$ & $\mathbf{21.05\pm53.49}$ \\
\bottomrule
\end{tabular}
\end{table}
%\FloatBarrier

\begin{table}[!htb]
\caption{\textcolor{black}{Patch-level model complexity for the proposed routed nnU-Net branches and representative alternative backbones. FLOPs are patch-level estimates, not direct case-level runtime measurements. Because the input patch size differs across models, FLOPs are not strictly comparable and indicate only order-of-magnitude differences.} %Patch-level model complexity for the proposed routed nnU-Net branches and representative alternative backbones. FLOPs are reported as patch-level estimates and are therefore not direct case-level runtime measurements.
}
\label{tab:model_complexity}
\centering
\tiny
\setlength{\tabcolsep}{1.1pt}
\renewcommand{\arraystretch}{1.12}
\begin{tabular}{>{\raggedright\arraybackslash}p{2.7cm}c c c >{\raggedright\arraybackslash}p{3.5cm}}
\toprule
Model & Input patch & Params (M) & FLOPs (G) & Notes \\
\midrule
Ours (routed nnU-Net branches) & $4\times128\times160\times112$ & $30.79$ & $1177.7$ & Same PlainConvUNet backbone for the pre-only and RC-aware joint branches; one routed model is applied per case. \\ \\
Swin UNETR & $4\times128\times128\times128$ & $62.20$ & $656.0$ & Local \texttt{feature\_size=48} implementation with sliding-window inference; changing the number of output heads has negligible parameter impact. \\ \\
SegMamba & $4\times128\times128\times128$ & $67.42$ & $3191.1$ & Same backbone for pre-treatment and post-treatment settings; the small difference caused by output-channel number is negligible. \\
\bottomrule
\end{tabular}
\end{table}
%\FloatBarrier

\subsection*{Qualitative analysis}

\begin{figure}[!htbp]
\centering
\includegraphics[width=\textwidth]{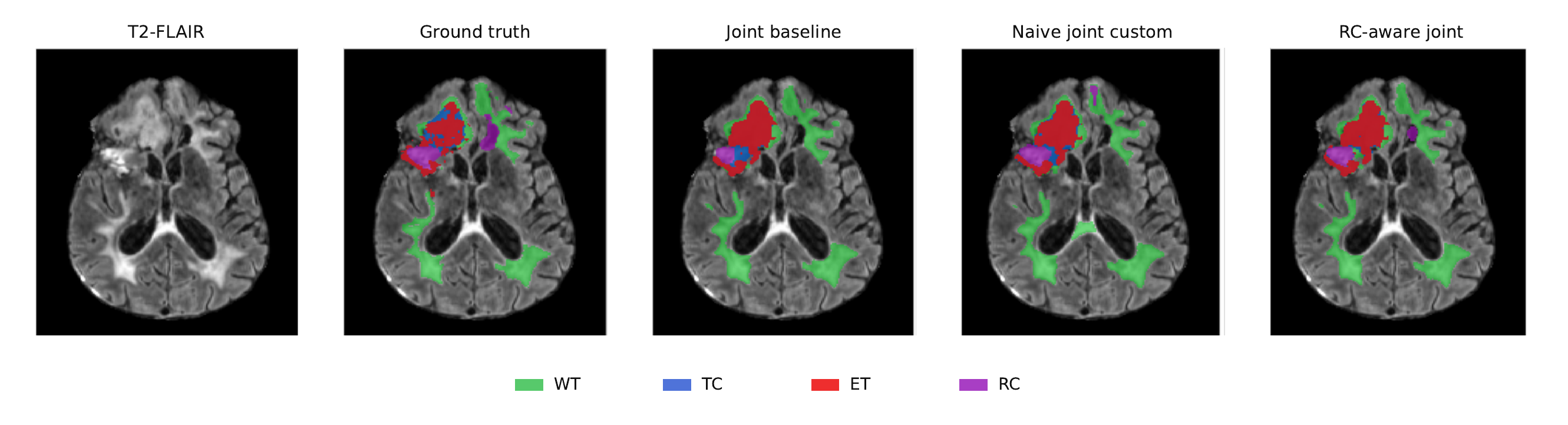}
\caption{Qualitative comparison on one post-treatment held-out test case. Columns: input
T2-FLAIR, ground truth, and predictions from Joint Baseline, Naive Joint
Custom, and RC-aware Joint. Overlays show the nested tumor regions (WT in green, TC in blue, and ET
in red) and the separate RC region (purple). Model outputs are converted
to the original BraTS-GLI categorical label space using the decoding
procedure described in  %Section~\ref{sec: method} 
the Methods section
before visualization. %Qualitative comparison of our nnU-Netv2-based ablation models on a representative post-treatment held-out test case. The columns show the input T2-FLAIR slice, ground-truth annotation, Joint Baseline prediction, Naive Joint Custom prediction, and RC-aware Joint prediction.  Colors are overlaid in nested BraTS-GLI evaluation regions: WT in green, TC in blue, ET in red, and RC in purple.  WT is defined as labels 1, 2, and 3; TC as labels 1 and 3; ET as label 3; and RC as label 4.  Predictions are visualized after mapping the model-output labels back to the original BraTS-GLI label space.
}
\label{fig:qualitative_post}
\end{figure}

Figure~\ref{fig:qualitative_post} shows one example from a post-treatment
held-out case (BraTS-GLI-02846-100), comparing the input T2-FLAIR slice
and ground-truth annotation with predictions from the Joint Baseline,
Naive Joint Custom, and RC-aware Joint models. The case was selected
because all four region outputs considered in our analysis are visible on the
displayed slice, allowing the nested region structure and RC behavior
to be inspected in a single example; it is intended only as one qualitative example, and predictions vary
across the test set. The figure is intended as a qualitative
illustration rather than a further statistical comparison.

The example highlights the setting targeted by RC-aware joint
supervision. Both joint-training alternatives largely miss the resection
cavity at its true location and place spurious RC voxels within the
tumor region, whereas the RC-aware joint model recovers the cavity in
the correct position while preserving the nested WT, TC, and ET
structure around the post-surgical region. This is consistent with the
higher RC lesion-wise Dice of the RC-aware joint model in
Table~\ref{tab:local_lwdice}.

%Figure~\ref{fig:qualitative_post} shows a representative post-treatment held-out test set case (BraTS-GLI-02846-100).  The figure compares the input T2-FLAIR slice, ground-truth annotation, and predictions from the joint baseline, naive joint custom-loss model, and RC-aware joint model.  This case was selected for visualization because all four post-treatment evaluation regions are visible on the displayed slice, allowing the nested region representation and RC behavior to be inspected in a single example. The figure is intended as a qualitative illustration rather than an additional statistical comparison. This case is representative of the trend in the quantitative results
%rather than a best case; regions and predictions vary across the test
%set. This example illustrates the post-treatment setting targeted by the proposed RC-aware joint supervision: compared with the two joint-training alternatives, the RC-aware joint model provides a more localized delineation of the resection cavity while maintaining the nested WT, TC, and ET structure around the post-surgical region.

%\FloatBarrier

% =========================================================

\section*{Discussion}
This paper presents a practical solution to BraTS-GLI 2025 based on an
RC-aware joint objective and treatment-status-guided routing. The results indicate that the
benefit of the proposed framework does not arise from a custom loss
alone; rather, the ablation study supports modeling pre-treatment and
post-treatment segmentation as a structured problem with
treatment-status-dependent imaging appearances and target definitions. Using this
design, our method achieved second place in the official BraTS-GLI 2025 testing phase.

The official top-ranked results show a clear trade-off between
performance and inference complexity. The first-place method achieved
the best reported test performance using a three-model ensemble with
on-the-fly GliGAN augmentation, and the two methods tied for third place
followed different routes, including a customized architecture and a
toolkit/post-processing-oriented solution. Our method achieved the
second-best official rank using treatment-status-guided routing, in which each
case was processed by a single specialized model. The small gap in the descriptive mean Dice and NSD@1.0 scores across
the four displayed regions between the first-place method and our
method suggests that task-structure-aware modeling can be competitive
without ensemble-based inference.

The patch-level complexity analysis provides additional context. The
proposed routed nnU-Net branches use fewer parameters than the evaluated
Swin UNETR and SegMamba variants and require substantially fewer
patch-level operations than SegMamba, although Swin UNETR has lower
patch-level FLOPs, in part because our branches use a larger input patch
($128\times160\times112$ vs.\ $128\times128\times128$). Notably, the
routed branches achieved this with roughly half the parameters of Swin
UNETR and SegMamba (30.8\,M vs.\ 62.2\,M and 67.4\,M) while obtaining the
best values in most regions and metrics on the internal held-out set,
showing that strong performance was achieved without increasing the
parameter count relative to the evaluated alternative backbones. Direct case-level runtime comparisons
would require matched sliding-window settings, hardware, and
implementation details.

The local ablation study provides diagnostic evidence for the final
design but should be interpreted in the context of the challenge
workflow. The routed model combination was chosen during the official
evaluation phase, whereas the internal split was constructed after the
challenge for controlled analysis. On this internal split, the custom
loss improved the post-only setting but did not consistently improve the
pre-only setting. This suggests that lesion- and boundary-aware losses
can be helpful in the more difficult post-treatment setting, but should
not be interpreted as universally superior across all data regimes. For
pre-treatment cases, the official online validation leaderboard
supported the routed pre-only custom-loss branch, whereas the
post-challenge internal split did not reproduce a uniform advantage over
the pre-only baseline. We therefore interpret the pre-treatment branch
as part of the challenge-phase routed solution rather than as definitive
evidence that the custom loss is universally better for all
pre-treatment splits.

The most direct ablation evidence supports the complete RC-aware joint
training configuration. Naive joint custom-loss training performed
worst on the post-treatment held-out test set, showing that simply
pooling the two cohorts and applying the single-cohort custom loss was
insufficient. The proposed joint model, which combines BCE,
boundary-weighted BCE, and empty-target-aware Dice handling, improved
post-treatment lesion-wise Dice, NSD@1.0, and HD95 over both joint
comparators. Because its Dice formulation and component weights differ
from those of the comparators, these results support the complete
configuration rather than establishing the independent causal effect of
the empty-target penalty. The strong RC performance on the official
test set is nevertheless consistent with the suitability of this
configuration for pooled pre-treatment and post-treatment training.

\textcolor{black}{ A natural question is whether routing is necessary. On the internal split, the RC-aware joint model performed comparably to the pre-only branch on pre-treatment cases
(Tables~\ref{tab:local_lwdice}--\ref{tab:local_lwhd95}), suggesting that
a single RC-aware joint model applied to all cases could be a simpler
alternative that also removes the need for treatment-status metadata at
inference. The routed configuration was retained here because it was
selected during the challenge phase.}  

From a statistical design perspective, the patient-level split is also
important. Because some patients contribute scans at multiple
timepoints, splitting by scan could leak patient-specific information
from training to validation. Grouping all scans from the same patient
into the same subset provides a more credible post-challenge estimate of
generalization, especially for models trained across treatment statuses.

\subsection*{Limitations}
\label{sec:limitations}
This study has several limitations. First, the post-challenge local
ablation uses a single patient-level split rather than repeated
train/validation resampling or cross-validation. Second, the local
ablation was performed after the challenge and cannot fully reconstruct
the official validation-leaderboard model-selection process. Third, the
local ablation relies on the BraTS-GLI lesion-wise evaluator and may not
exactly reproduce the full official blind testing-phase ranking
procedure. Fourth, the lesion-aware loss is a differentiable surrogate
rather than the exact connected-component matching procedure used by the
official challenge evaluation. Fifth, the treatment-status-guided inference
strategy requires each test case to be assigned to either the
pre-treatment or post-treatment group before prediction; in the BraTS-GLI 2025  data this is available from the case identifier and
challenge-provided data organization, whereas datasets without such
metadata would require an additional treatment-status classifier or a manual
check. Finally, although our method uses no model-level probability
averaging, voting, or multi-architecture ensembling, it uses two
specialized treatment-status-specific models and the nnU-Netv2 default
sliding-window inference with test-time mirroring; this should be
interpreted as treatment-status-guided routing rather than as a single universal
model.

\subsection*{Future work}
Future work could use repeated patient-level resampling or external
validation to quantify the stability of the observed differences beyond
the current split. A dedicated RC false-positive analysis on pre-treatment cases, together
with a component-wise ablation of the empty-target penalty and loss
weights, would help separate the contribution of individual elements of
the joint objective from that of the complete training configuration.
Another direction is a unified architecture with a treatment-status embedding
or gating mechanism, allowing treatment-status-specific behavior to be learned
within a single network. Finally, architecture-level comparisons could
be expanded to additional backbones and repeated splits to better
separate architecture effects from split-specific variability.

\section*{Conclusions}
We presented RC-aware nnU-Netv2, a simple yet effective framework for pre-
and post-treatment glioma segmentation using multimodal MRI, which achieved second
place in BraTS-GLI 2025
with performance comparable to that of the first-place method. Our framework combines lesion- and boundary-aware
single-cohort training, an RC-aware joint objective for pooled
pre-treatment and post-treatment data, and treatment-status-guided
routing without multi-model ensembling. On the official blind test set,
the method achieved arithmetic mean lesion-wise Dice and NSD@1.0 scores
of 0.8306 and 0.8334 across the four displayed regions (ET, RC, TC, and
WT), respectively. Post-challenge evaluation showed that the
RC-aware joint model improved post-treatment lesion-wise Dice, NSD@1.0,
and HD95 over the joint baseline and naive joint custom-loss training,
although the experiments compare complete training configurations
rather than individual loss components. Overall, the results demonstrate
that the proposed framework provides a competitive approach with a low active parameter count per case to pre- and post-treatment glioma
segmentation, with each case processed by one routed model rather than
an ensemble.
% =========================================================

% =========================================================

% =========================================================

%\section*{Statements and Declarations}

\section*{List of abbreviations}\label{sec:List_of_abbreviations}

\begin{tabular}{@{}ll@{}}
\toprule
\textbf{Abbreviation} & \textbf{Full Term} \\
\midrule
3D & Three-dimensional \\
BCE & Binary cross-entropy\\
BraTS & Brain Tumor Segmentation\\
BraTS-GLI 2025 & BraTS 2025 Lighthouse Challenge Task 1: Glioma Segmentation \\
&on Pre- and Post-treatment MRI\\
CI & Confidence interval\\
ET& Enhancing tumor or enhancing tissue\\
FLAIR & Fluid-attenuated inversion recovery\\
FLOPs & Floating-point operations \\
HD95 & 95th-percentile Hausdorff distance\\
LW & Lesion-wise\\
MIST & Medical Imaging Segmentation Toolkit\\
MONAI & Medical Open Network for Artificial Intelligence\\
MRI & Magnetic resonance imaging\\
N/A & Not applicable\\
NETC & Non-enhancing tumor core\\
NSD & Normalized surface Dice\\
NSD@1.0 & Normalized surface Dice at a 1.0 mm tolerance\\
RC & Resection cavity\\
SD & Standard deviation\\
SNFH & Surrounding non-enhancing T2-FLAIR hyperintensity\\
T1 & T1-weighted imaging\\
T1Gd & Post-contrast T1-weighted imaging\\
T2 & T2-weighted imaging\\
T2-FLAIR & T2-weighted fluid-attenuated inversion recovery\\
TC& Tumor core\\
WT&Whole tumor\\
\bottomrule
\end{tabular}

\subsection*{Author contributions} 
All authors contributed to the study conception and design. L.Q. performed material preparation, data collection, and analysis. L.Q., Z.C., and H.S. wrote the manuscript, and all authors commented on earlier versions. All authors read and approved the final manuscript.

\subsection*{Funding}
Dr. Ziqi Chen’s work was partially supported by the National Natural Science Foundation of China (grant numbers 12671307, 12271167, and 72331005). Dr. Hai Shu’s work was partially supported by NYU GPH Research Support Grant.

\subsection*{Data Availability}
The data used in this study were released for BraTS 2025 Lighthouse Challenge Task~1 (BraTS-GLI 2025) and are available to registered participants through the official Synapse challenge
platform (https://www.synapse.org/Synapse:syn64153130/) under the applicable registration and
data-use terms. Official validation and blind-test labels were not released to participating teams.

\subsection*{Code Availability}
Code for the customized trainers, inference wrapper, and reproducibility documentation will be
released upon acceptance of the manuscript.

\subsection*{Ethics, Consent to Participate, and Consent to Publish Declarations}
Not applicable.

\subsection*{Competing Interests}
The authors declare no competing interests.

\iffalse
\subsection*{Ethics Approval}
This study
used existing de-identified data and involved no direct contact with
participants; therefore, no additional institutional ethics approval
was required.
\fi

\bibliography{bibliography}

\end{document}